\documentclass[12pt]{article}

\catcode`\@=11
\@addtoreset{equation}{section}

\global\arraycolsep=2pt

\usepackage{mathrsfs,amsbsy,amssymb,latexsym,amsfonts,amsmath,cite}
\usepackage{graphicx,color}
\usepackage{mathtools}

\usepackage[bulletsep]{collref}
\collectsep{$\blacklozenge$\,}

\allowdisplaybreaks

\newcommand{\cF}{{\mathcal F}}

\newcommand{\tJ}{{\tilde J}}  

\newcommand{\al}{\alpha}
\newcommand{\be}{\beta}
\newcommand{\ga}{\gamma}
\newcommand{\de}{\delta}
\newcommand{\ep}{\epsilon}

\newcommand{\la}{\lambda}

\newcommand{\om}{\omega}

\newcommand{\cE}{\mathcal E}
\newcommand{\cZ}{\mathcal Z}
\newcommand{\cL}{\mathcal L}

\newcommand{\Tr}{{\rm Tr}}

\newcommand{\alg}[1]{\mathfrak{#1}}
\newcommand{\el}{\nonumber}
\newcommand{\nln}{\nonumber\\}

\newcommand{\pexp}{{\rm P}\!\exp}

\newcommand{\tE}{\tilde{\mathcal E}}
\newcommand{\tZ}{\tilde{\mathcal Z}}
\newcommand{\tL}{\tilde{\mathcal L}}

\begin{document}

\begin{flushright}
\parbox{4cm}
{KUNS-2534}
\end{flushright}

\vspace*{1.5cm}

\begin{center} 
{\Large\bf Yang-Baxter sigma models based on the CYBE} 
\vspace*{1.5cm}\\ 
{\large 
Takuya Matsumoto$^{\dagger}$\footnote{
E-mail:~takuya.matsumoto@math.nagoya-u.ac.jp} 
and Kentaroh Yoshida$^{\ast}$\footnote{
E-mail:~kyoshida@gauge.scphys.kyoto-u.ac.jp}
} 
\end{center}
\vspace*{0.25cm}
\begin{center}
$^{\dagger}${\it Institute for Advanced Research and 
Department of Mathematics, \\ Nagoya University, 
Nagoya 464-8602, Japan.} 
\vspace*{0.25cm}\\ 
$^{\ast}${\it Department of Physics, Kyoto University, \\ 
Kyoto 606-8502, Japan.} 
\end{center}
\vspace{1cm}

\begin{abstract}
It is known that Yang-Baxter sigma models 
provide a systematic way to study integrable deformations
of both principal chiral models and symmetric coset sigma models. 
In the original proposal and its subsequent development, 
the deformations have been characterized by classical $r$-matrices 
satisfying the modified classical Yang-Baxter equation (mCYBE). 
In this article, we propose the Yang-Baxter sigma models 
based on the classical Yang-Baxter equations (CYBE) rather than the mCYBE. 
This generalization enables us to utilize various kinds of solutions of 
the CYBE to classify integrable deformations. 
In particular, it is straightforward to realize 
partial deformations of the target space without loss of the integrability 
of the parent theory.  
\end{abstract}

\vspace*{1cm}


\setcounter{footnote}{0}
\setcounter{page}{0}
\thispagestyle{empty}

\newpage

\tableofcontents

\section{Introduction}

In the recent years, it has been an intriguing subject to consider  
integrable deformations of type IIB superstring  
on the AdS$_5\times$S$^5$ background (which is often abbreviated to 
the AdS$_5\times$S$^5$ superstring). 
Such directions would be important to reveal the underlying common 
dynamics of the AdS/CFT correspondence \cite{M}. 

\medskip 

In the duality between the AdS$_5\times$S$^5$ superstring and 
the ${\cal N}=4$ super Yang-Mills theory in four dimensions, 
the integrability is recognized to play important roles 
(For a comprehensive review, see \cite{review}). 
On the string-theory side, the classical integrability was argued  
by constructing the Lax connection on the world-sheet \cite{BPR}. 
The existence of the Lax pair is based on the $\mathbb{Z}_4$-grading property of the 
supercoset of the AdS$_5\times$S$^5$ superstring action \cite{MT}. 
On the gauge-theory side, it was shown that the one-loop planar dilatation operator 
for composite operators of adjoint scalar fields   
is mapped to an $SO(6)$ integrable spin-chain Hamiltonian \cite{MZ}. 
This was only the tip of an iceberg of the whole integrable structure behind 
the AdS/CFT correspondence. 

\medskip 

While, the underlying fundamental principle of the AdS/CFT duality 
has not been completely understood yet. 
In order to gain further insights for this issue, we expect that deformations 
of the integrable structures appearing in the AdS/CFT 
would be a clue to reveal it. 
For instance, introducing some deformation parameters enables us to take
various limits of them. 
By doing so, we are able to discuss a family of dualities 
with keeping the integrable structure. 
One of such attempts is a $q$-deformation of the dynamical spin-chain 
model, which possesses a centrally extended $\alg{su}(2|2)$ symmetry 
\cite{BK}. The whole symmetry algebra is naturally enhanced to an infinite 
dimensional quantum affine algebra \cite{BGM}. 

\medskip 

From the AdS/CFT point of view, it is desirable to figure out similar integrable 
deformations of the AdS$_5\times$S$^5$ string action. 
However, in comparison to 
deformations of a quantum integrable model 
with a manifest algebraic symmetry, it is not so obvious to realize  
those of an integrable classical filed theory.  
This difficulty has been quite alleviated by a notion
of the {\it Yang-Baxter sigma models} introduced by Klimcik, 
which are integrable deformations of principal chiral models
\cite{K-YB1,K-YB2,K-biYB}. 
In this description, integrable deformations are characterized by 
R-operators satisfying the {\it modified} classical Yang-Baxter equations (mCYBE).
After these remarkable works, Delduc, Magro and Vicedo generalized  
the formulation to purely bosonic coset sigma models
\cite{DMV} and finally succeeded to an integrable deformation of 
the AdS$_5\times$S$^5$ string action \cite{DMV2,DMV3}. 
This deformation could be regarded as a classical analogue of $q$-deformation
of Lie algebras \cite{Drinfeld1,Drinfeld2,Jimbo} because it is based on the 
Drinfeld-Jimbo type classical $r$-matrix and, indeed, the Poisson symmetry algebra
is a quantum group associated with the superconformal algebra,  
${\cal U}_q(\alg{psu}(2,2|4))$ \cite{DMV3}.    
The resulting deformed metric and NS-NS two-form are explicitly derived in \cite{ABF}.

\medskip 

On the other hand, an alternative integrable deformation of AdS$_5\times$S$^5$
has been proposed in \cite{KMY-Jor}, where deformations are 
characterized by classical $r$-matrices satisfying the classical Yang-Baxter
equation (CYBE) rather than the mCYBE\footnote{
Another type of deformation is also proposed in \cite{HMS-def,S-lambda}.}.  
Hence, this is a natural generalization of Yang-Baxter deformations of the 
AdS$_5\times$S$^5$ superstring \cite{DMV2,DMV3}. 
There are two important aspects of the deformations based on the CYBE 
in comparison to the mCYBE. 
Firstly, the CYBE allows various kinds of the skew-symmetric constant solutions, 
which are convenient in studying the deformations.  
Secondary, it is possible to consider partial deformations of the target space. 
The two properties of the CYBE are significant in terms of applicability 
because the present formulation provides a systematic way to investigate 
a large class of integrable deformations by using solutions of the CYBE. 

\medskip 

In fact, as remarkable examples of deformations of CYBE-type, 
the Lunin-Maldacena-Frolov backgrounds \cite{LM,Frolov} 
and the gravity duals of non-commutative gauge theories \cite{HI,MR}
have been successfully recovered \cite{LM-MY,MR-MY}. 
The classical integrability of these backgrounds automatically follows 
from the construction. 
Based on these observations, it seems likely that the space of solutions 
of the CYBE may be identified with the moduli space of a certain class of solutions of 
type IIB supergravity. It is referred as to the {\it gravity/CYBE correspondence} 
(For a short review, see \cite{MY-review}).  
In fact, the full supergravity solution \cite{SUGRA-KMY} associated with a deformation 
argued in \cite{KMY-Jor} is also obtained by a chain of string dualities 
such as TsT-transformations and S-dualities \cite{MY-TsT}. 
The deformation technique is also applicable to a Sasaki-Einstein manifold $T^{1,1}$\,, 
because it is described as a coset \cite{CMY}.  
The resulting deformed background agrees with the one obtained in \cite{LM,CO}. 

\medskip 

In this article, inspired by these developments in the AdS$_5\times$S$^5$ 
superstring, we will show that integrable deformations based on the CYBE 
work also for bosonic principal chiral models. 
More precisely, our main statement is the following. 
Let an R-operator $R$ be a solution of the CYBE. 
Then the classical action of the deformed Yang-Baxter sigma model is given by
\begin{align}
S=-\frac{1}{2}\int^\infty_{-\infty}\!\!\!d\tau\int^{2\pi}_0\!\!\!d\sigma~
\Tr\left((g^{-1}\partial_-g)\frac{1}{1-\eta R}(g^{-1}\partial_+g)\right)\,. 
\nonumber 
\end{align}
Here $g$ is a $G$-valued function on the string world-sheet with the coordinates $\tau$ and $\sigma$\,, 
where $G$ is a Lie group and 
$\partial_\pm=\partial_\tau\pm \partial_\sigma$
are derivatives with respect to the light-cone coordinates. 
The parameter $\eta$ measures the associated deformation. 
Note that the model is reduced to a principal chiral model
when $\eta=0$\,. 
Then, this deformed model has the following Lax pair;  
\begin{align}
\cL_\pm(\la)
=\frac{1}{1\pm\la}\left(1-\frac{\la\eta R}{1\pm \eta R}\right) 
(g^{-1}\partial_\pm g) \,, 
\label{Lax-intro} 
\end{align}
where $\la\in \mathbb{C}$ is a spectral parameter. 
Thus, the resulting deformed model is also classically integrable. 
A similar generalization is also possible for bosonic coset 
sigma models, as explained in the main part of this article. 

\medskip 

This article is organized as follows. 
In Sec.\ \ref{sec:2}, we generalize Yang-Baxter sigma models
and define deformed principal chiral models based on the CYBE. 
The Lax pair is also presented. 
In Sec.\ \ref{sec:3}, we argue a similar generalization for coset sigma models. 
In Sec.\ \ref{sec:multi-para}, we explain a  
multi-parameter generalization of CYBE-type. 
Sec.\ \ref{sec:4} is devoted to conclusion and discussion. 
App.\ \ref{app:eom} explains in detail a derivation of equation of motion
of the deformed coset sigma models. 
We also present some examples of mCYBE-type deformations 
in App.\ \ref{App:example}.

\section{Integrable deformations of principal chiral models} 
\label{sec:2}

In this section we shall discuss integrable deformations of principal chiral models.  
In Subsec.\ \ref{subsec:2.1}, 
after recalling the definition of the Yang-Baxter
sigma models based on the modified classical
Yang-Baxter equation (mCYBE) \cite{K-YB1,K-YB2},  
we show that the formulation can naturally be generalized 
for the standard classical Yang-Baxter equation (CYBE). 
Then the Lax pair is explicitly presented. 
Finally we demonstrate an example in Subsec.\ \ref{subsec:2.2}. 

\subsection{Yang-Baxter sigma models} 
\label{subsec:2.1}

\paragraph{Definition of the models}

Let $G$ be a Lie group and $\alg{g}$ be the associated Lie algebra. 
The action of the Yang-Baxter sigma models introduced by 
\cite{K-YB1,K-YB2,K-biYB} is given by 
\begin{align}
S=-\frac{1}{2}(\ga^{\al\be}-\ep^{\al\be})
\int^\infty_{-\infty}\!\!\!d\tau\int^{2\pi}_0\!\!\!d\sigma~
\Tr\left(A_\al\frac{1}{1-\eta R}A_\be \right)\,.
\label{action1}
\end{align}
Here $\xi^{\alpha}=(\xi^{\tau},\xi^{\sigma}) = (\tau,\sigma)$ 
are coordinates of the two-dimensional world-sheet. We will work with the flat metric 
$\ga^{\al\be}={\rm diag}(-1,1)$ in the conformal gauge. 
Then $\ep^{\al\be}$ is the skew-symmetric tensor
normalized as $\ep^{\tau\sigma}=-\ep^{\sigma\tau}=1$\,. 
The left-invariant one-form $A_\al=g^{-1}\partial_\al g$ is written in terms of 
$g(\tau, \sigma)\in G$ and $A_{\alpha}$ is a $\alg{g}$-valued function. 
The trace is computed over the fundamental representation of $\alg{g}$\,. 
A constant parameter $\eta$ measures a deformation of the model. 
When $\eta=0$\,, the action \eqref{action1} is nothing but that of the 
$G$-principal chiral models.

\medskip 

An important ingredient is a classical $r$-matrix denoted by $R$\,, 
which is an ${\mathbb R}$-linear operator $R:\alg{g}\to \alg{g}$ and satisfies 
the (modified) classical Yang-Baxter equations ((m)CYBE); 
\begin{align}
[R(X),R(Y)]-R([R(X),Y]+[X,R(Y)])=\omega [X,Y]  
\qquad \text{with} \qquad \om=\pm1, 0 \,,  
\label{CYBE}
\end{align}
where $X,Y\in \alg{g}$\,. 
When $\om=1$ (or $-1$), the $R$-operator satisfying \eqref{CYBE}
is called {\it non-split} (or {\it split}) type (respectively). 
In particular, when $\om=0$\,, the equation \eqref{CYBE} is reduced to 
the classical Yang-Baxter equation (CYBE). 
We assume that the $R$-operator 
does not depend on the spectral parameter and it 
is skew-symmetric; 
\begin{align}
\Tr(R(X)Y)=-\Tr(XR(Y)) \qquad \text{for}\qquad 
X,Y\in \alg{g}\,. 
\label{skew-R}
\end{align}

\medskip 

Note that a classical $r$-matrix in the tensorial notation is associated with 
an R-operator by tracing out the second entry, 
\begin{align}
R(X)=\Tr_2[r_{12}(1\otimes X)]\equiv \sum_i 
\bigl(a_i \Tr(b_i X)- b_i \Tr(a_i X)\bigr)
\qquad \text{for}\qquad X\in\alg{g} \,,  
\label{linear-R}
\end{align}
where the $r$-matrix is denoted symbolically as 
\begin{align}
r_{12}=\sum_i a_i\wedge b_i\equiv 
\sum_i \bigl(a_i\otimes b_i-b_i\otimes a_i\bigr)
\qquad \text{with}\qquad a_i\,, b_i\in \alg{g}\,. 
\end{align}

\medskip

For later convenience, we introduce the light-cone expressions of $A_{\alpha}$ like 
\begin{align}
A_\pm=A_\tau\pm A_\sigma\,. 
\label{lccoord}
\end{align}
With these notations, the Lagrangian of the action \eqref{action1} 
is recast into a simple form;  
\begin{align}
L=\frac{1}{2}\Tr\left(A_- J_+ \right)=\frac{1}{2}\Tr\left(A_+ J_- \right) 
\qquad \text{where} \qquad 
J_\pm:=\frac{1}{1\mp \eta R}A_\pm \,. 
\label{action2}
\end{align}

\paragraph{Equation of motion}

To obtain the equation of motion, 
let us take a variation of the Lagrangian \eqref{action2}\,. 
Defining a variation of $g\in G$ as 
$\de g=g \ep $
with an infinitesimal parameter $\ep$\,, the following relation is derived: 
\begin{align}
\de A_\al=\partial_\al \ep+[A_\al,\ep]\,. 
\end{align}
Then, the variation of \eqref{action2} is evaluated as  
$\de L=-\Tr(\cE \ep)$ with $\cE$ defined by 
\begin{align}
{\cal E}:=\partial_+J_-+\partial_-J_+-\eta([R(J_+),J_-]+[J_+,R(J_-)])\,. 
\end{align}
Thus, the equation of motion turns out to be $\cE=0$\,. 

\paragraph{Zero-curvature condition}

The next task is to rewrite the zero-curvature condition of 
$A_\pm=g^{-1}\partial_\pm g$ in terms of the deformed current $J_\pm$\,. 
For this purpose, let us introduce the following quantity; 
\begin{align}
\cZ:=\partial_+A_--\partial_-A_++[A_+,A_-]\,. 
\label{ZA}
\end{align}
By definition of $A_{\alpha}$\,, the zero-curvature condition is 
nothing but $\cZ=0$\,. 
Plugging $A_\pm=(1\mp \eta R)J_\pm$ 
with the above definition \eqref{ZA}\,, we obtain the following expression, 
\begin{align}
{\cal Z}&=\partial_+J_--\partial_-J_+-\eta([R(J_+),J_-]-[J_+,R(J_-)]) \el\\
&\quad +[J_+,J_-]-\eta^2\, {\rm YBE}(J_+,J_-)+\eta R({\cal E})
\end{align}
where the left-hand-side of the CYBE \eqref{CYBE} is denoted as  
\begin{align}
{\rm YBE}(X,Y):=[R(X),R(Y)]-R([R(X),Y]+[X,R(Y)])\,.   
\end{align}
When the $R$-operator is a solution of the (m)CYBE \eqref{CYBE},   
the quantity $\cZ$ becomes 
\begin{align}
{\cal Z}=\partial_+J_--\partial_-J_+-\eta([R(J_+),J_-]-[J_+,R(J_-)]) 
+(1-\eta^2\om)[J_+,J_-]+\eta R({\cal E})\,. 
\end{align}

\paragraph{Lax pair}
A novel feature of the Yang-Baxter sigma models \eqref{action1} is that 
there exists the Lax pair,  
\begin{align}
\cL_\pm(\la)= \left(\frac{1\mp \eta^2\om \la}{1\pm\la}\mp \eta R\right)J_\pm 
=\frac{1}{1\pm\la}\left(1\mp\frac{\eta\la(\eta\om\pm R)}{1\pm \eta R}\right)A_\pm
\label{YB-Lax}
\end{align}
with a spectral parameter $\la\in\mathbb{C}$\,. 
Indeed, the equation of motion $\cE=0$ and 
the zero-curvature condition $\cZ=0$ are equivalent to the flatness
condition of the Lax pair $\cL_\pm(\la)$\,; 
\begin{align}
\partial_+\cL_-(\la)-\partial_-\cL_+(\la)+[\cL_+(\la),\cL_-(\la)]=0\,. 
\label{flatness}
\end{align}
Thus, the models defined in \eqref{action1} are classically integrable 
in the sense of kinematical integrability. 

\medskip 

It should be emphasized that the Lax pair exists not only for 
$\om=1$ \cite{K-YB1,K-YB2}
but also for $\om=-1$ ({\it split}-type) and $\om=0$ (CYBE)\,.  
In particular, when the R-operator satisfies the CYBE with $\om=0$\,, 
the Lax pair in \eqref{YB-Lax} turns out to be 
\begin{align}
\cL_\pm(\la)= \left(\frac{1}{1\pm\la}\mp \eta R\right)J_\pm 
=\frac{1}{1\pm\la}\left(1-\frac{\la \eta R}{1\pm \eta R}\right)A_\pm\,. 
\label{YB-Lax-CYBE}
\end{align}
This is nothing but the Lax connection given in \eqref{Lax-intro}. 
Note that, when $\eta=0$\,, it reduces to the well-known Lax pair \cite{MZ-lax}
of principal chiral models, 
\begin{align}
\cL_\pm(\la)=\frac{A_\pm}{1\pm\la}\,. 
\end{align}

\medskip 

Let us prove the relation \eqref{flatness} is equivalent with 
the equation of motion $\cE=0$ and 
the zero-curvature condition $\cZ=0$ by a constructive way. 
In order to do this, we adopt the following ansatz (see also \cite{K-YB1,K-YB2}); 
\begin{align}
\cL_\pm(\la)= \left(\frac{F\pm G\la}{1\pm\la}\mp \eta R\right)J_\pm\,.
\label{L-ansatz}
\end{align}
Here it is supposed that the unknown variables $F$ and $G$ do not depend on
neither spectral parameter $\la$ nor the world-sheet coordinates ($\tau,\sigma$)\,. 
Under the ansatz \eqref{L-ansatz}\,, the flatness condition \eqref{flatness} 
can be rewritten as 
\footnote{
We are grateful to Jun-ichi Sakamoto for pointing out
some mistakes here. }
\begin{align}
0&=\partial_+\cL_-(\la)-\partial_-\cL_+(\la)+[\cL_+(\la),\cL_-(\la)] \nln 
&=\frac{1}{1-\la^2}
\Bigl[F\cZ+(F-1)\bigl((F+\eta^2\om)[J_+,J_-]-\eta R(\cE)\bigr)\Bigr]
+\frac{\la}{1-\la^2}(F-G)\cE \nln 
&-\frac{\la^2}{1-\la^2}
\Bigl[G\cZ+(G-1)\bigl((G+\eta^2\om)[J_+,J_-]-\eta R(\cE)\bigr)\Bigr]\,.  
\end{align}
Provided that the equation of motion $\cE=0$ and the zero-curvature condition 
$\cZ=0$ hold, 
the necessary conditions for the flatness of the Lax connection are
%the possible values of $F$ and $G$ are obtained as follows:  
\begin{align}
(F,G)=(1,1)\,,~(1,-\eta^2\om)\,,~(-\eta^2\om,1)\,,~(-\eta^2\om,-\eta^2\om)\,.
\end{align}
Inversely, to recover $\cE=\cZ=0$ from the  the flatness of the Lax pair,
it requires $F\neq G$ in addition to the above conditions
\footnote{Noting that, when $F=G=1$\,, 
the ansatz \eqref{L-ansatz} is nothing but $A_\pm$ itself.}\,. 
Therefore, the relevant choices for the Lax pair turn out to be 
\begin{align}
(F,G)=(1,-\eta^2\om)\quad \text{or}\quad(-\eta^2\om,1) \,.
\end{align}
Note that the associated two Lax pairs are related each other by $\la\to 1/\la$\,. 
Thus, taking $F=1$ and $G=-\eta^2\om$\,, 
we have the Lax pair presented in \eqref{YB-Lax}. 
This completes the proof.

\subsection{Example: 3D Schr\"odinger sigma models}
\label{subsec:2.2}

Let us see an example of the deformed principal chiral models in \eqref{action1}.  
This is a deformation of $SL(2;\mathbb{R})\simeq$ AdS$_3$ based on 
an $r$-matrix satisfying the CYBE with $\om=0$ rather than the mCYBE
\footnote{  
For an example of the mCYBE-type deformation, see App.\ \ref{App:example-PCM}.}.   
This model is  defined on a three-dimensional Schr\"odinger 
spacetime \cite{IKOP,Son,BMc} and hence it is often called 3D 
{\it Schr\"odinger sigma model}.
The classical integrable structure of this model has been discussed 
in \cite{KY-Sch,KMY-3DJordanian}.

\medskip 

Let $E,F,H$ be the generators of $\alg{sl}(2;\mathbb{R})$ satisfying the relations, 
\begin{align}
[F,E]=H\,, \qquad [H,E]=E\,, \qquad [H,F]=-F\,. 
\end{align}
With th generators, it is easy to see that the following classical $r$-matrix,  
\begin{align}
r_{12}=H\wedge F=H\otimes F-F\otimes H\,,  
\label{Jor-r}
\end{align}
satisfies the CYBE in \eqref{CYBE} with $\om=0$\,. 
We refer the $r$-matrix of this type as to {\it Jordanian-type} 
because it has non-zero Cartan charges.  
Then, we will find that the $r$-matrix (\ref{Jor-r}) yields sigma models defined on 
the 3D Schr\"odinger geometry; 
\begin{align}
ds^2=\frac{-2dx^+dx^-+dz^2}{z^2}-\frac{\eta^2(dx^+)^2}{4z^4} \,, 
\label{3D-sch}
\end{align}
where $\eta$ is a deformation parameter. 
Note that the scalar curvature of this metric is equal to  
that of AdS$_3$\,, namely, 
\begin{align}
R=-6\,. 
\end{align}

\medskip 

In order to derive the deformed metric \eqref{3D-sch} from the Jordanian 
$r$-matrix \eqref{Jor-r}, 
we use the fundamental representation of $\alg{sl}(2;\mathbb{R})$\,,
\begin{equation}
E=\frac{1}{\sqrt{2}}\begin{pmatrix} 0&1\\0&0 \end{pmatrix}\,,
\qquad 
F=\frac{-1}{\sqrt{2}}\begin{pmatrix} 0&0\\1&0 \end{pmatrix}\,,
\qquad 
H=\frac{1}{2}\begin{pmatrix} 1&0\\0&-1 \end{pmatrix}\,. 
\end{equation}
Using these matrices, we parametrize an $SL(2;\mathbb{R})$ group element as 
\begin{align}
g= {\rm e}^{2x^+ E} {\rm e}^{2(\log z) H} {\rm e}^{2x^- F}\,. 
\end{align}
Then, the left-invariant one-form $A_\pm=g^{-1}\partial_\pm g $ is evaluated as 
\begin{align}
A_\al=A_\al^E F +  A_\al^F E+A_\al^H H
\end{align}
with the coefficients 
\begin{align}
A_\pm^E&=2\partial_\pm x^- 
+\frac{4x^-(x^-\partial_\pm x^+
-z\partial_\pm z)}{z^2}\,, 
\nln 
A_\pm^F&=\frac{2\partial_\pm x^+}{z^2}\,,
\nln
A_\pm^H&=\frac{2(z\partial_\pm z-2x^-\partial_\pm x^+)}{z^2} \,. 
\end{align}

\medskip 

Next, to compute the deformed current $J_\pm$\,, it is necessary to figure out the 
action of the linear R-operator associated with the Jordanian 
$r$-matrix \eqref{Jor-r}. 
Taking the trace over the second sites,  
the R-operator and its transformation law turn out to be  
\begin{align}
&R(X)=\Tr_2[r_{12}(1\otimes X)] \qquad \text{for} \qquad 
X \in\alg{sl}(2;\mathbb{R})\,, \nln 
\Longrightarrow\quad 
&R(E)=-\frac{1}{2}H\,, \qquad 
R(H)=-\frac{1}{2}F\,, \qquad 
R(F)=0\,. \label{trans}
\end{align}
In particular, note that the R-operator is nilpotent; $R^3=0$\,. 
This property enables us to compute the current explicitly as follows; 
\begin{align}
J_\pm &= \frac{1}{1\mp \eta R} A_\pm  
= (1\pm \eta R+\eta^2 R^2) A_\pm \nln 
&=\left(A_\pm^E \mp\frac{\eta}{2} A^H_\pm+\frac{\eta^2}{4}A^F_\pm\right)F+
\left(A_\pm^H \mp\frac{\eta}{2} A^F_\pm\right) H +A_\pm^FE \,. 
\end{align}

\medskip 

Finally, one can rewrite the Lagrangian as 
\begin{align}
L&=\frac{1}{2}\Tr[A_-J_+] \nln
&=-\ga^{\al\be} \left(\frac{-2\partial_\al x^+ \partial_\be x^-
+\partial_\al z \partial_\be z}{z^2}
-\frac{\eta^2 \partial_\al x^+ \partial_\be x^+}{4z^4} \right)
-\eta\ep^{\al\be} \partial_\al x^+ 
\partial_\be \left(\frac{1}{2z^2}\right)\,.  
\end{align}
The last term coupled with the anti-symmetric tensor is total derivative, 
and hence it can be omitted. 
Indeed, this is nothing but a non-linear sigma model defined on 
the 3D Schr\"odinger spacetime \eqref{3D-sch}. 
The classical integrability of this model follows automatically 
because the Lax pair is explicitly obtained 
by plugging (\ref{trans}) with the expression (\ref{YB-Lax-CYBE}).

\medskip 

It would be interesting to try to reveal the relation between the above construction 
and the coset construction argued in \cite{SYY}. 
The symmetric two-form discussed in \cite{SYY} would possibly be related to 
the classical $r$-matrix (\ref{Jor-r}).

\section{Integrable deformations of coset sigma models}
\label{sec:3}

In the next, we will extend the previous argument on deformed principal chiral models 
to purely bosonic sigma models defined on symmetric cosets. 
This is a natural generalization of \cite{DMV}
\footnote{For an example of the mCYBE-type deformation
of symmetric cosets, see App.\ \ref{App:example-coset}.}.

\subsection{Yang-Baxter deformations of symmetric cosets}

\paragraph{Symmetric cosets}

Recall first the definition of symmetric cosets. 
Let $G$ be a Lie group and $H$ be a subgroup of $G$\,.
The associated Lie algebras of $G$ and $H$ are denoted by 
$\alg{g}$ and $\alg{g}^{(0)}$\,, respectively. 
The Lie algebra $\alg{g}$ is a direct sum of $\alg{g}^{(0)}$ and its 
complementary space $\alg{g}^{(1)}$ as a vector space;   
\begin{align}
\alg{g}=\alg{g}^{(0)}\oplus\alg{g}^{(1)}\,. 
\end{align}
Then, the homogeneous space $G/H$ is called 
{\it symmetric space} if $\alg{g}^{(0)}$ and $\alg{g}^{(1)}$ satisfy the following 
$\mathbb{Z}_2$-grading property, 
\begin{align}
[\alg{g}^{(0)},\alg{g}^{(0)}]\subset \alg{g}^{(0)}\,, \qquad 
[\alg{g}^{(0)},\alg{g}^{(1)}]\subset \alg{g}^{(1)}\,, \qquad 
[\alg{g}^{(1)},\alg{g}^{(1)}]\subset \alg{g}^{(0)}\,.  
\label{sym-pair}
\end{align}
The pair ($\alg{g}^{(0)}$\,, $\alg{g}^{(1)}$) satisfying  
the above property is often referred as to {\it symmetric pair}.  
It is convenient to introduce a projector to the subspace 
$\alg{g}^{(1)}$ by 
\begin{align}
P:~~\alg{g}\longrightarrow \alg{g}^{(1)}\,. 
\label{proj}
\end{align}

\paragraph{Definition of the coset models}

Yang-Baxter deformations of symmetric coset sigma models 
have been introduced in \cite{DMV} and the action is given by 
\begin{align}
S=-\frac{1}{2}(\ga^{\al\be}-\ep^{\al\be})
\int^\infty_{-\infty}\!\!\!d\tau\int^{2\pi}_0\!\!\!d\sigma~
\Tr\left(A_\al P\frac{1}{1-\eta R_g\circ P}A_\be \right)\,. 
\label{coset-action}
\end{align}
The coset projector $P$ is given in \eqref{proj} and 
the dressed R-operator $R_g$ is defined by   
\begin{align}
R_g(X)\equiv g^{-1}R(gXg^{-1})g \qquad \text{with}\qquad 
g\in G\,.  
\label{dressed-R}
\end{align}
Here $\eta$ is a deformation parameter. 
The action is reduced to that of the undeformed coset sigma model when $\eta = 0$\,. 

\medskip 

By using the light-cone notation \eqref{lccoord}\,, 
the Lagrangian can be rewritten as 
\begin{align}
L=\frac{1}{2}\Tr\bigl(A_- P(\tJ_+) \bigr)
=\frac{1}{2}\Tr\bigl(A_+ P(\tJ_-) \bigr)\,,  
\label{coset-action2}
\end{align}
where we have introduced the deformed current,  
\begin{align}
\tilde J_\pm:=\frac{1}{1\mp \eta P\circ R_g}A_\pm \,. 
\label{def-cur}
\end{align}

\paragraph{Equation of motion}

The equation of motion of the model \eqref{coset-action2} is given by 
$\tE=0$ where 
\begin{align}
\tE:=\partial_+P(\tJ_-)+\partial_-P(\tJ_+)+[\tJ_+,P(\tJ_-)]+[\tJ_-,P(\tJ_+)]\,. 
\label{eom-coset}
\end{align}
For the detail of the derivation, see App.\ \ref{app:eom}.

\paragraph{Zero-curvature condition}

By definition of the left-invariant one-form $A_\al=g^{-1}\partial_\al g$\,, 
it satisfies the zero-curvature condition $\tZ=0$\,, where  
\begin{align}
\tZ:=\partial_+A_--\partial_-A_++[A_+,A_-]\,. 
\end{align}
Plugging the relation $A_\pm=(1\mp \eta P\circ R_g)\tJ_\pm$ with the 
above expression, one can recast it into the following form,
\begin{align}
\tZ=\partial_+\tJ_--\partial_-\tJ_++[\tJ_+,\tJ_-]
+\eta R_g(\tE)+\eta^2\, {\rm YBE}_g(P(\tJ_+),P(\tJ_-))\,. 
\end{align}
Here we have used the bookkeeping notation;  
\begin{align}
{\rm YBE}_g(X,Y):=[R_g(X),R_g(Y)]-R([R_g(X),Y]+[X,R_g(Y)])\,.   
\end{align}
When the R-operator satisfies the (m)CYBE;
\begin{align}
{\rm YBE}_g(X,Y)=\om [X,Y]\,,    
\end{align}
the expression of $\tZ$ further reduces to 
\begin{align}
\tZ=\partial_+\tJ_--\partial_-\tJ_++[\tJ_+,\tJ_-]
+\eta R_g(\tE)+\eta^2\om [P(\tJ_+),P(\tJ_-)]\,. 
\label{zero-rewrite}
\end{align}

\paragraph{Lax pair} 

We are ready to construct the Lax pair of the model \eqref{coset-action2}. 
Indeed, both the equation of motion $\tE=0$ 
and the zero-curvature condition $\tZ=0$ 
are equivalent to the flatness condition of the Lax pair, 
\begin{align}
\partial_+\tL_-(\la)-\partial_-\tL_+(\la)+[\tL_+(\la),\tL_-(\la)]=0\,, 
\label{flatness-coset}
\end{align}
where $\tL_\pm(\la)$ is defined as  
\begin{align}
\tL_\pm(\la) \equiv \tJ_\pm^{(0)}+\la^{\pm1} \sqrt{1+\eta^2\om} \tJ_\pm^{(1)} \,. 
\label{lax-coset}
\end{align}

\medskip 

It should be noted that the above Lax pair is flat not only for 
$\om=1$ (split type) \cite{DMV} but also  
$\om=-1$ (non-split type) and 
$\om=0$ (CYBE-type). 
In particular, when the R-operator satisfies the CYBE with $\om=0$\,, 
the Lax pair \eqref{lax-coset} becomes 
\begin{align}
\tL_\pm(\la)=\tJ_\pm^{(0)}+\la^{\pm1} \tJ_\pm^{(1)} \,.  
\label{lax-coset-CYBE}
\end{align}
Interestingly, this is of the same form with the Lax pair of the 
undeformed coset sigma model, up to a formal replacement 
$A_\pm \to \tJ_\pm$\,. 

\medskip 

To find out the Lax connection \eqref{lax-coset}\,,   
we start with the following ansatz,   
\begin{align}
\tL_\pm(\la)=\tJ_\pm^{(0)}+\la^{\pm1} \tilde G \tJ_\pm^{(1)} \,.
\label{coset-ans}
\end{align}
Here we suppose that the unknown factor $\tilde G$ does not depend on neither 
the spectral parameter $\la$ nor the world-sheet coordinates ($\tau,\sigma$)\,. 
Under this ansatz, the flatness condition \eqref{flatness-coset} 
can be rewritten as  
\begin{align}
0&=\partial_+\tL_-(\la)-\partial_-\tL_+(\la)+[\tL_+(\la),\tL_-(\la)] \nln 
&=-\tilde G (\partial_-\tJ_+^{(1)}-[\tJ_+^{(1)},\tJ_-^{(0)}])\la \nln
&\quad +\tilde G (\partial_+\tJ_-^{(1)}+[\tJ_+^{(0)},\tJ_-^{(1)}])\la^{-1}  \nln 
&\quad +\partial_+\tJ_-^{(0)}-\partial_-\tJ_+^{(0)}+[\tJ_+^{(0)},\tJ_-^{(0)}]
+\tilde G^2[\tJ_+^{(1)},\tJ_-^{(1)}]\,. 
\label{comp1}
\end{align}

\medskip 

On the other hand, due to the symmetric property \eqref{sym-pair}\,, 
the equation of motion $\tE=0$ and the zero-curvature condition 
$\tZ=0$ are equivalent with the following set of the three equations; 
\begin{align}
0&=\partial_-\tJ_+^{(1)}-[\tJ_+^{(1)},\tJ_-^{(0)}]\,,  \\
0&=\partial_+\tJ_-^{(1)}+[\tJ_+^{(0)},\tJ_-^{(1)}]\,, \\ 
0&=\partial_+\tJ_-^{(0)}-\partial_-\tJ_+^{(0)}+[\tJ_+^{(0)},\tJ_-^{(0)}]
+(1+\eta^2\om)[\tJ_+^{(1)},\tJ_-^{(1)}]\,. 
\end{align}
Comparing these relations with \eqref{comp1}\,, one can find the following relation,  
\begin{align}
\tilde G=\pm \sqrt{1+\eta^2\om}\,. 
\end{align}
The overall sign does not matter because it can be absorbed by 
the redefinition of the spectral parameter $\la$\,. 
When the plus signature is adopted, the ansatz \eqref{coset-ans}
agrees with the Lax pair \eqref{lax-coset}\,. 

\subsection{Twist operator} 
\label{subsec:twist} 

From the expression \eqref{zero-rewrite}\,, it should be noted that 
the deformed current $\tilde J_\pm$ is also flat if the equation of motion is satisfied 
$\tE=0$ and the R-operator is a solution of the CYBE with $\omega=0$\,; 
\begin{align}
\cZ=\partial_+\tJ_--\partial_-\tJ_++[\tJ_+,\tJ_-]=0\,. 
\label{onshell-flat}
\end{align}
In other words, the current $\tJ_\pm$ is {\it on-shell flat current} 
regarding the CYBE-type deformation. 
Hence, one may expect that there exists a group element $\cF\in G$\,, 
which we call {\it twist operator}, such that the deformed current is expressed as 
\begin{align}
\tJ_\pm=\tilde g^{-1}\partial_\pm \tilde g \qquad \text{with} \qquad 
\tilde g\equiv \cF^{-1} g\,. 
\label{J-twist}
\end{align}
With this notation, the flatness of $J_\pm$ is obvious. 
In the following, we will concretely construct such a twist operator. 

\medskip 

For this purpose, we suppose that the sigma model \eqref{coset-action}
is defined on an infinitely extended world-sheet parametrized by 
$\sigma\in (-\infty,+\infty)$\,, instead of a cylinder; 
\begin{align}
S=-\frac{1}{2}(\ga^{\al\be}-\ep^{\al\be})
\int^\infty_{-\infty}\!\!\!d\tau\int^{\infty}_{-\infty}\!\!\!d\sigma~
\Tr\left(A_\al P\frac{1}{1-\eta R_g\circ P}A_\be \right)\,. 
\end{align}
Then the field $g$ and the current $A_\al$ are also supposed to obey
the following boundary conditions on the world-sheet,  
\begin{align}
g(\sigma=\pm \infty)=\text{const.}\qquad \Longrightarrow\qquad 
A_\al(\sigma=\pm \infty)=0\,. 
\label{bc}
\end{align}

\medskip 

Let us next consider the gauge transformation of $J_\pm$ defined by 
\begin{align}
\tJ_\pm^g\equiv g\tJ_\pm g^{-1}-\partial_\pm gg^{-1} 
\qquad\Longleftrightarrow\qquad 
\partial_\pm+\tJ_\pm^g\equiv g(\partial_\pm+\tJ_\pm)g^{-1}\,,  
\label{gauge-trf}
\end{align}
which is explicitly calculated as 
\begin{align}
\tJ^g_\pm =g(\tJ_\pm-A_\pm)g^{-1}
=g\Bigl(\frac{\pm\eta R_g\circ P}{1\mp\eta R_g\circ P}A_\pm \Bigr)g^{-1}
=\pm \eta R(gP(\tJ_\pm)g^{-1})\,.  
\end{align}
Because the current $\tJ_\pm$ is {\it on-shell} flat current as we have seen in 
\eqref{onshell-flat}, the gauge transformed current  $\tJ_\pm^g$ is also flat. 
By taking account of the boundary condition \eqref{bc}, this observation leads us to 
introduce the following twist operator by 
\begin{align}
\cF(\sigma,\tau)\equiv \pexp\Bigl[
-\int^\sigma_{-\infty}\!\!d\sigma\, \tJ^g_\sigma \Bigr]K\,,  
\end{align}
where $K\in G$ is a constant element and does not depend on 
the world-sheet variables $(\tau,\sigma)$\footnote{
The choice of a constant element $K$ would be crucial 
to find out the symmetry algebra of the deformed model. 
A suitable fixing of the matrix $K$ for the 3D Schr\"odinger sigma model 
has been argued in \cite{KMY-3DJordanian}.}. 
By the definition, it is easily shown that 
\begin{align}
\tJ^g_\pm=-\partial_\pm\cF \cF^{-1}\,. 
\end{align}
Plugging this expression with \eqref{gauge-trf}, one can obtain the following relation,  
\begin{align}
\tJ_\pm=-g^{-1}(\partial_\pm\cF \cF^{-1}) g + g^{-1}\partial_\pm g 
=(\cF^{-1}g)^{-1}\partial_\pm (\cF^{-1}g)\,. 
\end{align}
This is indeed the desired form in \eqref{J-twist}.

\section{Multi-parameter deformations based on the CYBE} 
\label{sec:multi-para}

One may expect a generalization of integrable deformations based on the CYBE
to the multi-parameter case. 
An easy way of doing this is to notice the following fact. 

\medskip 

Let $r^A_{12}$ and $r^B_{12}$ be solutions of the CYBE with $\om=0$\,. 
Suppose that they commute each other; $[r^A_{ij},r^B_{kl}]=0$ for $i,j,k,l\in\{1,2,3\}$\,.  
Then, a linear combination of the $r$-matrices defined as   
\begin{align}
r_{12}^{(\al,\be)}\equiv \al\, r^A_{12} + \be\, r^B_{12}\,,  \qquad (\al,\be\in\mathbb{C})\,, 
\end{align}
is also a solution of the CYBE. 
As a matter of course, the associated linear R-operator $R^{(\al,\be)}$ 
\eqref{linear-R} satisfies the CYBE with $\om=0$ in \eqref{CYBE}. 

\medskip 

Plugging the R-operator  $R^{(\al,\be)}$ 
with the action \eqref{action1} (or \eqref{coset-action}), 
one can obtain a two-parameter integrable deformation of 
principal chiral models (or coset sigma models, respectively)\footnote{
Apparently, there are three deformation parameters, $\al,\be$ and $\eta$\,, 
in \eqref{action1} or \eqref{coset-action}\,. 
In the present case, however, $\eta$ is regarded as the overall factor 
of the R-operator and can be absorbed by redefinition of $\al,\be$\,.
Hence, two of them are the independent deformation parameters.}.  
Repeating this step, it is easy to realize further multi-parameter deformations. 

\section{Conclusion and discussion}
\label{sec:4}

In this article, we have shown that the Yang-Baxter sigma models
introduced in \cite{K-YB1,K-YB2} can naturally be extended 
to the CYBE case. 
The deformed model is defined by \eqref{action1} and 
the Lax pair of the CYBE-type is obtained in \eqref{YB-Lax-CYBE}\,. 
We have also argued a generalization to symmetric coset sigma models 
as in \cite{DMV}. 
In this case, the action is given in 
\eqref{coset-action} and the Lax pair is given in \eqref{lax-coset-CYBE}\,. 

\medskip 

As mentioned in Introduction, these generalizations would be important 
from the viewpoint of applications because the CYBE has a wider class of the 
skew-symmetric constant solutions in general rather than the mCYBE. 
In particular, partial deformations of the target space manifestly 
preserve the classical integrability. 
Remarkably, multi-parameter generalizations are straightforward  
by following the technique explained in Sec.\ \ref{sec:multi-para}. 

\medskip 

As a future direction, it would be interesting to consider a similar generalization
for a two-parameter deformation of the principal chiral model, 
which is called {\it bi-Yang-Baxter sigma model} \cite{K-biYB}.  
In the recent progress, this formulation has been argued for  a superstring theory 
on an AdS$_3\times$S$^3\times$M$^4$ background \cite{H-biYB,HRT}.

\subsection*{Acknowledgments}

We are grateful to Io Kawaguchi for collaboration at the early stage of this work. 
We appreciate Gleb Arutyunov, Riccardo Borsato,
P.\ Marcos Crichigno, Marc Magro and Benoit Vicedo  
for valuable comments and discussions. 
This work is supported in part by JSPS Japan-Hungary 
Research Cooperative Program.

\appendix

\section*{Appendix}

\section{Derivation of the equation of motion \eqref{eom-coset}} 
\label{app:eom}

In this Appendix, we shall derive the equation of motion 
\eqref{eom-coset} for the deformed coset model. 

\medskip 

The variation of the undeformed current is computed as 
\begin{align}
\de A_\pm =\partial_\pm \ep +[A_\pm,\ep]
\qquad \text{under} \qquad \de g=g\ep\,. 
\end{align}
To evaluate the variation of the deformed current $\de\tJ_\pm$\,, 
we need some preparation. 
Firstly, by the definition of the dressed R-operator $R_g$ in \eqref{dressed-R}, 
one can derive the relation, 
\begin{align}
\de (R_g\circ P)(X) =(R_g\circ P)(\de X)+[(R_g\circ P)(X),\ep]-R_g([P(X),\ep])\,, 
\end{align}
where $X$ is an arbitrary field and $P$ is the coset projector defined in \eqref{proj}. 
Secondary, using this relation repeatedly, one can show that 
\begin{align}
\de \bigl((R_g\circ P)^n(X)\bigr) = (R_g\circ P)^n(\de X)
+\sum_{k=0}^{n-1}(R_g\circ P)^k[(R_g\circ P)^{n-k}(X),\ep] \nln 
-\sum_{k=0}^{n-1}(R_g\circ P)^k R_g([P\circ (R_g\circ P)^{n-1-k}(X),\ep])   
\end{align}
for a natural number $n$\,. 
Thirdly, multiplying $(\pm\eta)^n$ on both hand sides of the above relation
and summing up $n$ from $0$ to $\infty$\,, the following expression is obtained,  
\begin{align}
\de \left(\frac{1}{1\mp\eta R_g\circ P}X\right)
&=\frac{1}{1\mp\eta R_g\circ P}\nln
&\quad \times
\left(\de X
+\left[\frac{\pm\eta R_g\circ P}{1\mp\eta R_g\circ P}X,\ep\right]
\mp\eta R_g \left[P\frac{1}{1\mp\eta R_g\circ P}X,\ep\right]\right)\,. 
\end{align}
Finally, substituting $A_\pm$ for $X$\,, we have derived the variation of $\tJ_\pm$\,, 
\begin{align}
\de \tJ_\pm=\frac{1}{1\mp\eta R_g\circ P}\bigl(
\partial_\pm \ep +[\tJ_\pm,\ep]\mp \eta R_g [P(\tJ_\pm),\ep]\bigr)\,. 
\label{var-J}
\end{align}
Here we have used the definition of the deformed current in \eqref{def-cur}. 

\medskip 

We are now ready to evaluate the variation of the action in \eqref{coset-action}. 
Using the above formula \eqref{var-J}
and noting the skew-symmetric property of the R-operator \eqref{skew-R}, 
one can find the following relation; 
\begin{align}
\de L = -\frac{1}{2}\Tr \bigl[\bigl(
\partial_+P(\tJ_-)+\partial_-P(\tJ_+)+[\tJ_+,P(\tJ_-)]+[\tJ_-,P(\tJ_+)]\bigr)\ep\bigr]\,. 
\end{align}
This is nothing but the equation of motion $\tE=0$ with 
$\tE$ given in \eqref{eom-coset}. 

\section{Examples} 
\label{App:example} 

Here we present examples of deformations based on the mCYBE
for both principal chiral models and symmetric coset sigma models. 

\subsection{A squashed sigma model} 
\label{App:example-PCM} 

We consider a sigma model defined on a deformed S$^3$\,, called squashed three-sphere. 
This model is referred as to a {\it squashed sigma model}
or an {\it anisotropic principal chiral model} in \cite{Cherednik,FR,BFP}. 
The classical integrable structure has been studied in \cite{KY,KMY-QAA,Kame-AdS3,ORU}.

\medskip 

Let $T^\pm, T^3$ be the generators of $\alg{su}(2)$ satisfying the relations; 
\begin{align}
[T^3,T^\pm]=\pm i T^\pm \,, \qquad [T^+,T^-]= iT^3 \,. 
\end{align}
Now we consider the following classical $r$-matrix, 
\begin{align}
r_{12}=2 i T^+\wedge T^- =2i(T^+\otimes T^--T^-\otimes T^+)\,. 
\label{DJ-r}
\end{align}
We refer this type of $r$-matrix as to Drinfeld-Jimbo type. 
The associated R-operator satisfies the mCYBE \eqref{CYBE} with $\om=1$
(split-type). 
The resulting deformed spacetime turns out to be  
a squashed three-sphere; 
\begin{align}
ds^2=\frac{1}{4}\left(
d\theta^2 +\sin^2\theta d\phi^2 
+(1+\eta^2) (d\psi+\cos\theta d\phi)^2 \right) 
\label{sq-S3}
\end{align}
with a deformation parameter $\eta$\,. 
The scalar curvature is 
\begin{align}
R=6-2\eta^2\,. 
\end{align}

\medskip 

To derive the metric of the squashed S$^3$ in \eqref{sq-S3} from the $r$-matrix 
of Drinfeld-Jimbo type \eqref{DJ-r}, 
we introduce the fundamental representation of $\alg{su}(2)$ as follows; 
\begin{align}
&T^1=-\frac{i}{2}\left(
\begin{array}{cc}
 0 & 1 \\
 1 & 0 \\
\end{array}\right)\,,
\qquad 
T^2=-\frac{i}{2}\left(
\begin{array}{cc}
 0 & -i \\
 i & 0 \\
\end{array}\right)\,,
\qquad 
T^3=-\frac{i}{2}\left(
\begin{array}{cc}
 1 & 0 \\
 0 & -1 \\
\end{array}\right)\,. 
\end{align}
It is also convenient to introduce the light-cone notation, 
\begin{align}
&T^+=\frac{T^1-i T^2}{\sqrt{2}} 
=\frac{-i}{\sqrt{2}}\left(
\begin{array}{cc}
 0 & 1 \\
 0 & 0 \\
\end{array}\right)\,,
\qquad 
T^-=\frac{T^1+i T^2}{\sqrt{2}} 
=\frac{-i}{\sqrt{2}}
\left(
\begin{array}{cc}
0 & 0 \\
1 & 0 \\
\end{array}\right)\,. 
\end{align}
By using these generators, an SU(2) group element $g$ can be 
represented by 
\begin{align}
g= {\rm e}^{\phi T^3} {\rm e}^{\theta T^2} {\rm e}^{\psi T^3}\,. 
%=\left(
%\begin{array}{rr}
% e^{-\frac{i}{2} (\phi +\psi )} \cos \frac{\theta}{2}  
%& -e^{-\frac{i}{2}(\phi -\psi)} \sin\frac{\theta}{2}  \\
% e^{\frac{i}{2}(\phi -\psi )} \sin \frac{\theta}{2} 
%& e^{\frac{i}{2} (\phi +\psi )} \cos \frac{\theta}{2}  \\
%\end{array}
%\right)
%\in SU(2)\,, 
\end{align}
Then, the left-invariant current reads 
\begin{align}
A_\pm=g^{-1}\partial_\pm g 
=A^+_\pm T^-+A^-_\pm T^++A^3_\pm T^3
\end{align}
with the coefficients 
\begin{align}
A^-_\pm
&=\frac{i}{\sqrt{2}} e^{-i \psi } (\partial_\al \theta +i \partial_\al \phi  \sin\theta)\,, 
\nln
A^+_\pm
&=\frac{-i}{\sqrt{2}}e^{i \psi}( \partial_\al\theta -i \sin\theta\partial_\al \phi )\,,
\nln
A^3_\pm &=\partial_\al \psi+\cos\phi \partial_\al \theta\,. 
\end{align}

\medskip 

Let us next evaluate the deformed current $J_\pm$\,. 
Unlike the Jordanian $r$-matrix in Subsec.\ \ref{subsec:2.2}, 
the R-operator of Drinfeld-Jimbo type is not nilpotent but diagonally acts on 
the Chevalley-Serre generators;  
\begin{align}
&R(X)=\Tr_2[r_{12}(1\otimes X)] \qquad \text{for} \qquad 
X \in\alg{su}(2)\,, \nln 
\Longrightarrow\quad 
&R(T^\pm)=\mp iT^\pm \,, \qquad R(T^3)=0\,. 
\end{align}
By taking this into account, the deformed current is evaluated as follows; 
\begin{align}
J_\pm = \frac{1}{1\mp \eta R} A_\pm 
=\frac{1}{1\pm i\eta}A_\pm^- T^++\frac{1}{1\mp i\eta} A_\pm^+ T^-+A_\pm^3 T^3 \,.  
%&=\frac{i}{\sqrt{2}(1\pm i\eta)} e^{-i \psi } (\partial_\pm \theta 
%+i \partial_\pm \phi  \sin\theta)T^+
%-\frac{i }{\sqrt{2}(1\mp i\eta)}e^{i \psi}
%( \partial_\pm\theta -i \sin\theta\partial_\pm \phi )T^- \nln
%&\quad +(\partial_\pm \psi+\cos\phi \partial_\pm \theta  )T^3 
\end{align}

\medskip 

Finally, with the deformed current $J_\pm$\,, the 
Lagrangian is rewritten as \footnote{Here we have normalized the overall factor
in front of the action so that it agrees with \eqref{sq-S3}.}
\begin{align}
L&=-\frac{1+\eta^2}{2}\Tr[A_-J_+] =-\frac{1+\eta^2}{2}\Tr[A_+J_-]\nln
&=-\frac{\ga^{\al\be}}{4}\left(
\partial_\al \theta\partial_\be \theta+\sin^2\theta\partial_\al \phi\partial_\be \phi
+(1+\eta^2) (\partial_\al\psi+\cos\theta\partial_\al \phi )
 (\partial_\al\psi+\cos\theta\partial_\al \phi )\right) \nln 
&\quad +\frac{\ep^{\al\be}}{2}\,\eta\,\partial_\al\phi\, \partial_\be(\cos\theta)\,.   
\end{align}
Note that the last anti-symmetric term is total derivative and 
hence it can be dropped off.  
As a result, this is nothing but a sigma model 
on a squashed three-sphere \eqref{sq-S3}.

\subsection{A deformed sigma model on $SO(4)/SO(3)$}
\label{App:example-coset} 

The next is a typical example of deformed coset models based on the mCYBE. 
Here we consider a deformation of a symmetric coset representation of 
S$^3\simeq SO(4)/SO(3)$\,. 
Interestingly, the resulting background is different from the deformation of 
S$^3 \simeq SU(2)$\,. 
That is, the resulting deformed background depends on the representation of S$^3$\,. 

\medskip 

To describe the Lie algebra $\alg{so}(4)\simeq \alg{su}(2)\oplus \alg{su}(2)$\,, 
we prepare two sets of $\alg{su}(2)$ algebras generated by 
$A^i$ and $B^i$ with $i=1, 2, 3$ satisfying the relations; 
\begin{align}
[A^i,A^j]=\ep^{ij}{}_k A^k \,, \qquad [B^i,B^j]=\ep^{ij}{}_k B^k\,,
\qquad [A^i,B^j]=0\,.  
\end{align}
Here the structure constants $\ep^{ij}{}_k$ are totally anti-symmetric 
and normalized as $\ep^{12}{}_3=1$\,. 

\medskip 

We are concerned here with a classical $r$-matrix of the Drinfeld-Jimbo type, 
\begin{align}
r_{12}=i(A^+\wedge A^-+B^+\wedge B^-)\,.
\label{DJ-coset} 
\end{align}
Here $a\wedge b \equiv a\otimes b-b\otimes a$ and we have introduced 
the following notation, 
\begin{align}
A^\pm=-i A^1\pm A^2\,, \qquad B^\pm=-i B^1\pm B^2\,.  
\end{align}
It is easy to see that the associated $R$-operator is a solution of the 
mCYBE \eqref{CYBE} with $\om=1$\,. Thus it is of non-split type. 
The resulting deformed metric turns out to be 
\begin{align}
ds^2=\frac{1}{4}\left(\frac{d\theta^2-\eta^2 \sin^2\theta (d\phi-d\psi)^2/4}
{1+\eta^2\cos^2\frac{\theta}{2}}
+\sin^2\theta d\phi^2+(d\psi+\cos\theta\phi)^2\right)\,.  
\label{def-coset}
\end{align}
This background is not identical with the squashed three spheres \eqref{sq-S3}. 
In fact, the scalar curvature of the above metric is given by 
\begin{align}
R=6+\frac{2 \eta ^2 (1+\eta^2 \cos\theta \cos^2\frac{\theta }{2})}
{1+\eta^2 \cos^2\frac{\theta }{2}}\,. 
\end{align}
Hence the two backgrounds are not related to each other through 
a coordinate transformation. 

\medskip 

To derive the metric \eqref{def-coset} from the $r$-matrix \eqref{DJ-coset}, 
we need to recall the symmetric structure of 
$\alg{so}(4)=\alg{so}(4)^{(0)}\oplus\alg{so}(4)^{(1)}$\,. 
These subspaces are spanned by the following generators; 
\begin{align}
\alg{so}(4)^{(0)}&={\rm span}\{~J^i=A^i+B^i~|~i=1,2,3~\}\,, \nln 
\alg{so}(4)^{(1)}&={\rm span}\{~K^i=A^i-B^i~|~i=1,2,3~\}\,, 
\end{align}
respectively. Indeed, they enjoy the $\mathbb{Z}_2$-grading property \eqref{sym-pair}
as follows,  
\begin{align}
[J^i,J^j]=\ep^{ij}{}_k J^k \,, \qquad [K^i,J^j]=\ep^{ij}{}_k K^k\,,\qquad 
[K^i,K^j]=\ep^{ij}{}_k J^k \,. 
\end{align}
Then, the coset projector $P:\alg{so}(4) \rightarrow \alg{so}(4)^{(1)}$
in \eqref{proj} is defined as 
\begin{align}
P(X)=-K^1\Tr[K^1X]-K^2\Tr[K^2X]-K^3\Tr[K^3X] 
\qquad \text{for} \qquad X\in \alg{so}(4)\,. 
\end{align}
Here the trace can be computed on the $4\times4$ fundamental representation of 
$\alg{su}(2)\oplus\alg{su}(2)$\,, 
\begin{align}
&A^1=\tfrac{i}{2}
\begin{psmallmatrix}0&1&0&0\\1&0&0&0\\0&0&0&0\\0&0&0&0\end{psmallmatrix}\,, 
\qquad 
A^2=\tfrac{i}{2}
\begin{psmallmatrix}0&i&0&0\\-i&0&0&0\\0&0&0&0\\0&0&0&0\end{psmallmatrix}\,, 
\qquad 
A^3=\tfrac{i}{2}
\begin{psmallmatrix}1&0&0&0\\0&-1&0&0\\0&0&0&0\\0&0&0&0\end{psmallmatrix}\,, 
\nln 
&
B^1=\tfrac{i}{2}
\begin{psmallmatrix}0&0&0&0\\0&0&0&0\\0&0&0&1\\0&0&1&0\end{psmallmatrix}\,, 
\qquad 
B^2=\tfrac{i}{2}
\begin{psmallmatrix}0&0&0&0\\0&0&0&0\\0&0&0&i\\0&0&-i&0\end{psmallmatrix}\,, 
\qquad 
B^3=\tfrac{i}{2}
\begin{psmallmatrix}0&0&0&0\\0&0&0&0\\0&0&1&0\\0&0&0&-1\end{psmallmatrix}\,. 
\end{align}

\medskip 

In the next, a group element is parametrized as follows; 
\begin{align}
g={\rm e}^{\phi A^3} {\rm e}^{\theta A^2} {\rm e}^{\psi A^3}\in SU(2)\times SU(2)\,. 
\end{align}
Then, the left-invariant current $A_\pm=g^{-1}\partial_\pm g$ reads 
\begin{align}
A_\pm
&=(\sin\psi\partial_\pm\theta -\sin\theta \cos\psi \partial_\pm \phi )A^1 
+(\cos\psi\partial_\pm \theta + \sin\theta \sin\psi\partial_\pm\phi )A^2 \nln 
&\quad +(\partial_\pm \psi+ \cos\theta \partial_\pm \phi)A^3\,. 
\end{align}

\medskip 

To evaluate the deformed coset action in \eqref{coset-action}\,,  
we need to find the projected deformed current $P(\tJ_\pm)$
rather than the deformed current in \eqref{def-cur} itself. 
The current $P(\tJ_\pm)$ is determined by solving 
the following equations; 
\begin{align}
(1\mp\eta P\circ R_g )P(\tJ_\pm) =P(A_\pm)\,. 
\label{eq-J}
\end{align}
Note that these equations are obtained from the definition 
of the deformed current \eqref{def-cur}
by inverting the operator $(1\mp\eta R_g\circ P)$ and multiplying the 
projector $P$ from the left. 

\medskip 

Since the above equations \eqref{eq-J} are valued in $\alg{so}(4)^{(1)}$\,, 
there are three independent equations with respect to $K^1\,, K^2$ and $K^3$\,. 
Solving the three equations, we obtain the following expression, 
\begin{align}
P(\tJ_\pm)&=\frac{-1}{4(1+ \eta^2\cos^2\frac{\theta}{2})}
\Bigl[
\Bigl( ((2+\eta^2 \cos^2\tfrac{\theta}{2})\cos\psi 
\pm\eta \sin\psi )\sin\theta\partial_\pm \phi  
\nln 
&\quad 
+2 (\pm\eta \cos^2\tfrac{\theta }{2} \cos\psi-\sin\psi)\partial_\pm\theta
\pm\eta(\pm\eta \cos^2\tfrac{\theta }{2}\cos\psi-\sin\psi )
\sin\theta\partial_\pm \psi 
\Bigr)K_1
\nln 
&-\Bigl(
((2+\eta^2 \cos^2\tfrac{\theta }{2})\sin\psi
\mp \eta \cos\psi )\sin\theta\partial_\pm \phi 
\nln
&\quad 
+2 (\pm\eta \cos^2\tfrac{\theta }{2} \sin\psi +\cos \psi)\partial_\pm\theta 
\pm \eta (\pm\eta \cos^2\tfrac{\theta }{2} \sin\psi+\cos\psi )\sin\psi\partial_\pm\psi 
\Bigr)K_2
\nln 
&-\Bigl(  
((2+\eta^2) \cos\theta+\tfrac{\eta^2}{2}(1+\cos^2\theta ))\partial_\pm \phi 
\nln
&\quad 
\mp \eta   \sin\theta \partial_\pm\theta 
+2 (1+\eta^2 \cos^4\tfrac{\theta }{2}) \partial_\pm\psi 
\Bigr)K_3 \Bigr]\,. 
\end{align}
Finally, with the above expression of $P(J_\pm)$\,, 
the Lagrangian can be rewritten as 
\begin{align}
L&=-\Tr[A_+P(\tJ_-)]=-\Tr[A_-P(\tJ_+)]  \nln
&=-\frac{1}{4}\ga^{\al\be}
\biggl[\frac{\partial_\al \theta\partial_\be\theta
-\eta^2 \sin^2\theta (\partial_\al\phi-\partial_\al\psi)
(\partial_\be\phi-\partial_\be\psi)/4}
{1+\eta^2\cos^2\frac{\theta}{2}}
\nln 
&\qquad \quad +\sin^2\theta \partial_\al \phi \partial_\be\phi
+(\partial_\al\psi+\cos\theta\partial_\al\phi)
(\partial_\be\psi+\cos\theta\partial_\be\phi)\biggr]
\nln 
&\quad -\frac{\eta\sin\theta}
{4\left(1+\eta^2 \cos^2\tfrac{\theta}{2}\right)}\ep^{\al\be} 
\partial_\al \theta (\partial_\be\phi -\partial_\be\psi )\,.  
\end{align}
Note that the anti-symmetric two-form in the last line is total derivative 
and hence it can be ignored. Thus, this metric agrees with \eqref{def-coset}\,.


\begin{thebibliography}{99}

\bibitem{M}  
  J.~M.~Maldacena,
  ``The large N limit of superconformal field theories and supergravity,''
  Adv.\ Theor.\ Math.\ Phys.\  {\bf 2} (1998) 231
  [Int.\ J.\ Theor.\ Phys.\  {\bf 38} (1999) 1113]. 
  [arXiv:hep-th/9711200].

\bibitem{review}
  N.~Beisert {\it et al.},
  ``Review of AdS/CFT Integrability: An Overview,'' 
  Lett.\ Math.\ Phys.\ {\bf 99} (2012) 3 [arXiv:1012.3982 [hep-th]]. 

\bibitem{BPR}
  I.~Bena, J.~Polchinski and R.~Roiban,
  ``Hidden symmetries of the AdS$_5\times$S$^5$ superstring,''
  Phys.\ Rev.\ D {\bf 69} (2004) 046002
  [hep-th/0305116].

\bibitem{MT}
  R.~R.~Metsaev and A.~A.~Tseytlin,
  ``Type IIB superstring action in AdS$_5\times$S$^5$ background,''  
  Nucl.\ Phys.\ B {\bf 533} (1998) 109  [hep-th/9805028].  

\bibitem{MZ}
  J.~A.~Minahan and K.~Zarembo,
  ``The Bethe ansatz for N=4 superYang-Mills,''
  JHEP {\bf 0303} (2003) 013
  [hep-th/0212208].

  
\bibitem{BK}
  N.~Beisert and P.~Koroteev,
  ``Quantum Deformations of the One-Dimensional Hubbard Model,''
  J.\ Phys.\ A {\bf 41} (2008) 255204
  [arXiv:0802.0777 [hep-th]].


\bibitem{BGM}
  N.~Beisert, W.~Galleas and T.~Matsumoto,
  ``A Quantum Affine Algebra for the Deformed Hubbard Chain,''
  J.\ Phys.\ A {\bf 45} (2012) 365206
  [arXiv:1102.5700 [math-ph]].

\bibitem{K-YB1}
 C.~Klimcik,
 ``Yang-Baxter sigma models and dS/AdS T duality,''  
 JHEP {\bf 0212} (2002) 051  
 [hep-th/0210095]. 

\bibitem{K-YB2}
 C.~Klimcik,
 ``On integrability of the Yang-Baxter sigma-model,''  
 J.\ Math.\ Phys.\  {\bf 50} (2009) 043508  
 [arXiv:0802.3518 [hep-th]].  

\bibitem{K-biYB}
  C.~Klimcik,
  ``Integrability of the bi-Yang-Baxter sigma-model,''
  Lett.\ Math.\ Phys.\  {\bf 104} (2014) 1095
  [arXiv:1402.2105 [math-ph]].



\bibitem{DMV}
  F.~Delduc, M.~Magro and B.~Vicedo,
  ``On classical q-deformations of integrable sigma-models,''  
  JHEP {\bf 1311} (2013) 192  [arXiv:1308.3581 [hep-th]]. 
 
\bibitem{DMV2}
  F.~Delduc, M.~Magro and B.~Vicedo,
  ``An integrable deformation of the AdS$_5\times$S$^5$ superstring action,''  
 Phys.\ Rev.\ Lett.\  {\bf 112} (2014) 051601
  [arXiv:1309.5850 [hep-th]].

\bibitem{DMV3}
  F.~Delduc, M.~Magro and B.~Vicedo,
  ``Derivation of the action and symmetries of the $q$-deformed AdS$_5\times$S$^5$ superstring,''
  arXiv:1406.6286 [hep-th].  

\bibitem{Drinfeld1}
   V.~G.~Drinfel'd,
   ``Hopf algebras and the quantum Yang-Baxter equation,'' 
   Sov.\ Math.\ Dokl.\ {\bf 32} (1985) 254.  

 \bibitem{Drinfeld2}
  V.~G.~Drinfel'd,
   ``Quantum groups,''
   J.\ Sov.\ Math.\  {\bf 41} (1988) 898 
   [Zap.\ Nauchn.\ Semin.\  {\bf 155}, 18 (1986)].

 \bibitem{Jimbo}
   M.~Jimbo,
   ``A $q$ difference analog of $U(g)$ and the Yang-Baxter equation,''
   Lett.\ Math.\ Phys.\  {\bf 10} (1985) 63.        
  
 
\bibitem{ABF}
    G.~Arutyunov, R.~Borsato and S.~Frolov,
  ``S-matrix for strings on $\eta$-deformed AdS$_5\times$S$^5$,'' 
  JHEP {\bf 1404} (2014) 002 [arXiv:1312.3542 [hep-th]]. 


\bibitem{KMY-Jor}
  I.~Kawaguchi, T.~Matsumoto and K.~Yoshida,
  ``Jordanian deformations of the AdS$_5\times$S$^5$ superstring,''
  JHEP {\bf 1404} (2014) 153
  [arXiv:1401.4855 [hep-th]].

\bibitem{HMS-def}
  T.~J.~Hollowood, J.~L.~Miramontes and D.~M.~Schmidtt,
  ``An Integrable Deformation of the $AdS_5 \times S^5$ Superstring,''
  J.\ Phys.\ A {\bf 47} (2014) 49,  495402
  [arXiv:1409.1538 [hep-th]]; 
%\bibitem{HMS-def}
%  T.~J.~Hollowood, J.~L.~Miramontes and D.~M.~Schmidtt,
  ``Integrable Deformations of Strings on Symmetric Spaces,''
  JHEP {\bf 1411} (2014) 009
  [arXiv:1407.2840 [hep-th]].

\bibitem{S-lambda}
  K.~Sfetsos,
  ``Integrable interpolations: From exact CFTs to non-Abelian T-duals,''
  Nucl.\ Phys.\ B {\bf 880} (2014) 225
  [arXiv:1312.4560 [hep-th]].

\bibitem{LM}
 O.~Lunin and J.~M.~Maldacena,
  ``Deforming field theories with $U(1) \times U(1)$ global symmetry and their gravity duals,''  
JHEP {\bf 0505} (2005) 033  [hep-th/0502086].

\bibitem{Frolov}
  S.~Frolov,
  ``Lax pair for strings in Lunin-Maldacena background,''
  JHEP {\bf 0505} (2005) 069
  [hep-th/0503201].
  
  
\bibitem{HI}
  A.~Hashimoto and N.~Itzhaki,
  ``Noncommutative Yang-Mills and the AdS / CFT correspondence,''
  Phys.\ Lett.\ B {\bf 465} (1999) 142
  [hep-th/9907166].

\bibitem{MR}
  J.~M.~Maldacena and J.~G.~Russo,
  ``Large N limit of noncommutative gauge theories,''
  JHEP {\bf 9909} (1999) 025
  [hep-th/9908134].  

\bibitem{LM-MY}
  T.~Matsumoto and K.~Yoshida,
  ``Lunin-Maldacena backgrounds from the classical Yang-Baxter equation - towards the gravity/CYBE correspondence,''
  JHEP {\bf 1406} (2014) 135
  [arXiv:1404.1838 [hep-th]].    

\bibitem{MR-MY}  
 T.~Matsumoto and K.~Yoshida,
  ``Integrability of classical strings dual for noncommutative gauge theories,''
  JHEP {\bf 1406} (2014) 163 
  [arXiv:1404.3657 [hep-th]].    

\bibitem{MY-review}
  T.~Matsumoto and K.~Yoshida,
  ``Integrable deformations of the AdS$_5\times$S$^5$ superstring and the classical Yang-Baxter equation -- Towards the gravity/CYBE correspondence --,''
  arXiv:1410.0575 [hep-th].

\bibitem{SUGRA-KMY}
 I.~Kawaguchi, T.~Matsumoto and K.~Yoshida,
  ``A Jordanian deformation of AdS space in type IIB supergravity,'' 
  JHEP {\bf 1406} (2014) 146 [arXiv:1402.6147 [hep-th]].    

\bibitem{MY-TsT}
  T.~Matsumoto and K.~Yoshida,
  ``Yang-Baxter deformations and string dualities,''
  arXiv:1412.3658 [hep-th].

\bibitem{CMY}
P.~M.~Crichigno, T.~Matsumoto and K.~Yoshida,
  ``Deformations of $T^{1,1}$ as Yang-Baxter sigma models,''
  JHEP {\bf 1412} (2014) 085
  [arXiv:1406.2249 [hep-th]].

\bibitem{CO}
  A.~Catal-Ozer,
  ``Lunin-Maldacena deformations with three parameters,''
  JHEP {\bf 0602} (2006) 026
  [hep-th/0512290].

\bibitem{MZ-lax}
  V.~E.~Zakharov and A.~V.~Mikhailov,
  ``Relativistically Invariant Two-Dimensional Models in Field Theory Integrable by the Inverse Problem Technique. (In Russian),''
  Sov.\ Phys.\ JETP {\bf 47} (1978) 1017
   [Zh.\ Eksp.\ Teor.\ Fiz.\  {\bf 74} (1978) 1953].

\bibitem{IKOP}
  D.~Israel, C.~Kounnas, D.~Orlando and P.~M.~Petropoulos,
  ``Electric/magnetic deformations of S**3 and AdS(3), and geometric cosets,''
  Fortsch.\ Phys.\  {\bf 53} (2005) 73
  [hep-th/0405213].

\bibitem{Son}
  D.~T.~Son,
  ``Toward an AdS/cold atoms correspondence: A Geometric realization of the Schrodinger symmetry,''
  Phys.\ Rev.\ D {\bf 78} (2008) 046003
  [arXiv:0804.3972 [hep-th]].

\bibitem{BMc}
  K.~Balasubramanian and J.~McGreevy,
  ``Gravity duals for non-relativistic CFTs,''
  Phys.\ Rev.\ Lett.\  {\bf 101} (2008) 061601
  [arXiv:0804.4053 [hep-th]].

\bibitem{KY-Sch}
  I.~Kawaguchi and K.~Yoshida,
  ``Classical integrability of Schr\"odinger sigma models and $q$-deformed Poincare symmetry,''  
JHEP {\bf 1111} (2011) 094  [arXiv:1109.0872 [hep-th]]; 
%\bibitem{KY-exotic}
``Exotic symmetry and monodromy equivalence in Schr\"odinger sigma models,''  
JHEP {\bf 1302} (2013) 024  [arXiv:1209.4147 [hep-th]]. 

\bibitem{KMY-3DJordanian}
  I.~Kawaguchi, T.~Matsumoto and K.~Yoshida,
  ``Schroedinger sigma models and Jordanian twists,''
  JHEP {\bf 1308} (2013) 013
  [arXiv:1305.6556, arXiv:1305.6556 [hep-th]].

\bibitem{SYY}
  S.~Schafer-Nameki, M.~Yamazaki and K.~Yoshida,
  ``Coset Construction for Duals of Non-relativistic CFTs,''
  JHEP {\bf 0905} (2009) 038
  [arXiv:0903.4245 [hep-th]].         

\bibitem{H-biYB}
  B.~Hoare,
  ``Towards a two-parameter q-deformation of AdS$_3\times$S$^3\times$M$^4$ superstrings,''
  arXiv:1411.1266 [hep-th].

\bibitem{HRT}
  B.~Hoare, R.~Roiban and A.~A.~Tseytlin,
  ``On deformations of AdS$_n \times$S$^n$ supercosets,'' 
  JHEP {\bf 1406} (2014) 002 [arXiv:1403.5517 [hep-th]].  
 
\bibitem{Cherednik}
  I.~V.~Cherednik, 
  ``Relativistically Invariant Quasiclassical Limits Of Integrable
  Two-Dimensional Quantum Models,''
  Theor.\ Math.\ Phys.\  {\bf 47} (1981) 422
  [Teor.\ Mat.\ Fiz.\  {\bf 47} (1981) 225].

\bibitem{FR}
  L.~D.~Faddeev and N.~Y.~Reshetikhin,
  ``Integrability of the principal chiral field model in (1+1)-dimension,''
  Annals Phys.\  {\bf 167} (1986) 227.      

\bibitem{BFP}
  J.~Balog, P.~Forgacs and L.~Palla,
  ``A two-dimensional integrable axionic sigma model and T duality,''  
  Phys.\ Lett.\ B {\bf 484} (2000) 367  
  [hep-th/0004180].  

\bibitem{KY}
  I.~Kawaguchi and K.~Yoshida,
  ``Hidden Yangian symmetry in sigma model on squashed sphere,''
  JHEP {\bf 1011} (2010) 032. 
  [arXiv:1008.0776 [hep-th]]; 
% \bibitem{KYhybrid}
%  I.~Kawaguchi and K.~Yoshida,
  ``Hybrid classical integrability in squashed sigma models,''
  Phys.\ Lett.\ B\ {\bf 705} (2011) 251
  [arXiv:1107.3662 [hep-th]]; 
   ``Hybrid classical integrable structure of squashed sigma models: A short summary,''  
  J.\ Phys.\ Conf.\ Ser.\  {\bf 343} (2012) 012055 
  [arXiv:1110.6748 [hep-th]].    
  
\bibitem{KMY-QAA}
  I.~Kawaguchi, T.~Matsumoto and K.~Yoshida,
  ``The classical origin of quantum affine algebra in squashed sigma models,''  
  JHEP {\bf 1204} (2012) 115  [arXiv:1201.3058 [hep-th]]; 
%\bibitem{KMY-monodromy}
%  I.~Kawaguchi, T.~Matsumoto and K.~Yoshida,
  ``On the classical equivalence of monodromy matrices in squashed sigma model,''  
  JHEP {\bf 1206} (2012) 082  [arXiv:1203.3400 [hep-th]].

\bibitem{Kame-AdS3}
  T.~Kameyama and K.~Yoshida,
  ``String theories on warped AdS backgrounds and integrable deformations of spin chains,''
  JHEP {\bf 1305} (2013) 146
  [arXiv:1304.1286 [hep-th]].

\bibitem{ORU}
  D.~Orlando, S.~Reffert and L.~I.~Uruchurtu,
  ``Classical integrability of the squashed three-sphere, warped AdS3 and
  Schr$\ddot{\rm o}$dinger spacetime via T-Duality,''
  J.\ Phys.\ A  {\bf 44} (2011) 115401.
  [arXiv:1011.1771 [hep-th]]. 



\end{thebibliography}
\end{document}